\documentclass[useAMS,usenatbib,dcolumn,usegraphicx,twocolumn,10pt]{mnarxiv}
\usepackage{hyperref}
\usepackage{times}
\usepackage{placeins}
\usepackage{natbib}
\usepackage{ulem}
\newcommand{\ud}[1]{\mathrm{d}#1}

\newcommand{\br}[1]{\left(#1\right)}

\newcommand{\eqref}[1]{(\ref{#1})}

\newcommand{\kpc}{\mathrm{kpc}}

\newcommand{\msun}{\mathrm{M}_{\sun}}

\newcommand{\cut}{_{_{cut}}}
\newcommand{\rot}{_{_{rot}}}

\title[]{A possible influence of magnetic fields on the rotation of gas\\ in NGC~253}
\author[]{Joanna Ja{\l}ocha$^{1}$,
{{\L}ukasz Bratek$^{1}$},
{Jan P\c{e}kala$^{1}$}
{Marek Kutschera$^{2}$,}
\\
$^{1}$Institute of Nuclear Physics,
Polish Academy of Sciences, Radzikowskego 152, PL-31342 Krak\'{o}w, Poland\\
$^{2}$Institute of
Physics, Jagellonian University,  Reymonta 4, PL-30059 Krak{\'o}w, Poland}

\begin{document}
\date{\today}
\pagerange{\pageref{firstpage}--\pageref{lastpage}} \pubyear{2009}

\maketitle

\begin{abstract}
The magnetic fields that are present in the galaxy NGC 253 are
exceptionally strong. This means that they can influence the
rotation of matter and hence the mass-to-light ratio. In this
context, we address the issue of the presence of a non-baryonic
dark matter halo in this galaxy.
\medskip
\hrule
\begin{small}
\flushleft \textbf{The definitive version is available at\\
\url{http://onlinelibrary.wiley.com/doi/10.1111/j.1365-2966.2012.21967.x/abstract}}
\end{small}
\medskip
\hrule
\end{abstract}

\section{Introduction}
NGC 253 is a nearby late-type starburst spiral galaxy. Its
rotation curve, which has been measured by
\citet{1991AJ....101..456P} and \citet{1995AJ....110..199A},
increases with radial distance. However, another measurement by
\citep{1997ApJ...490..143B}, supported by
\citep{2011MNRAS.411...71H}, has indicated that the rotation
decreases for larger radii. This fall-off suggests that NGC 253
could be poor in non-baryonic dark matter. We calculate the
mass-to-light ratio by treating NGC 253 as a flattened disc-like
object, and we find that this ratio is low.

\newcommand{\muG}{\mathrm{\mu G}}

Several examples of other spiral galaxies have been described
consistently within the framework of the global disc model, which
suggests they could be disc-like objects without a massive
nonbaryonic dark matter halo
\citep{2008ApJ...679..373J,2010MNRAS.406.2805J,2010MNRAS.407.1689J}.
Moreover, magnetic fields can help to reduce the missing mass
problem. As shown by
\citep{1992Natur.360..652B,2004AcPPB..35.2493K,2008LNEA....3...83B},
magnetic fields can influence the rotation of partially ionized
gas. In the disc of NGC 253, there are exceptionally strong
magnetic fields present, which reach $18 \muG$
\citep{2009AN....330.1028H,2009A&A...506.1123H,2011A&A...535A..79H,2009A&A...494..563H}
These coexist with diffused warm ionized gas, extending out to
large radii \citep{2011MNRAS.411...71H}. We attempt to estimate
the influence of such fields on the rotation and the mass-to-light
ratio in this galaxy. Here, we consider only regular (large-scale)
fields. Given that small-scale (turbulent) fields are usually at
least as strong as the regular field, this (analytically
necessary) omission could have a significant effect.

\section{Surface mass  density and mass-to-light ratio} 
\begin{figure} \centering
\includegraphics[width=0.5\textwidth]{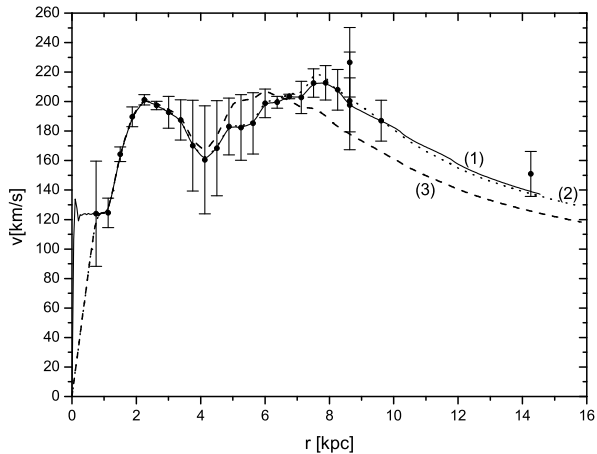}
\caption{\label{fig:rot} Rotation curve of the spiral galaxy NGC
253. The full circles with error bars represent the measurements
taken from  \citep{2011MNRAS.411...71H}. The solid line (labelled
1) denotes the model rotation curve obtained in the global disc
model. The dotted line (labelled 2) denotes a small modification
of the rotation curve (used in the text to show that a small
change in the rotation can greatly influence the mass-to-light
ratio). The dashed line (labelled 3) denotes a test rotation curve
discussed in the text (the rotation was lowered at larger radii so
that the mass-to-light ratio is reduced and decreases with large
galactocentric distances). } \end{figure}

We use the
rotation data from \citep{2011MNRAS.411...71H} shown in figure
\ref{fig:rot}.
 The last measurement
point located at $r\rot=14.3\,\kpc$ is separated from other
rotation data extending out to radius $r\cut=9.6\,\kpc$. For
numerical reasons, we treat the points with $r<r\cut$ as the
rotation curve, while the last point at $r\rot$ serves as a
control point. The surface brightness measurements in the K filter
extend out to large radii with the outermost point at
$r_K=11.8\,\kpc$ \citep{2003AJ....125..525J}. Since $r_K>r\cut$
these measurements can be used to constrain the surface mass
density for $r>r\cut$.

To find a global surface mass density, both the rotation and the
brightness data are needed. For lower radii ($r<r\cut$), the
rotation curve is used, while for $r\in (r\cut,r_K)$ the surface
mass density and the surface brightness are assumed to be
proportional to each other. In effect, a global rotation curve
corresponding to the (still unknown) global surface mass density
should agree (within error limits) with the rotation at $r\rot$.
To meet these requirements, the global surface mass density was
found with the help of the iteration method described by
\citep{2008ApJ...679..373J}. In the current context, this method
allows us to obtain a surface mass density that agrees both with
the rotation data for $r<r\cut$ and with the brightness
measurements for $r\in(r\cut,r_K)$ (in the original situation,
neutral hydrogen measurements were used for $r>r\cut$).

\begin{figure} \centering
\includegraphics[width=0.5\textwidth]{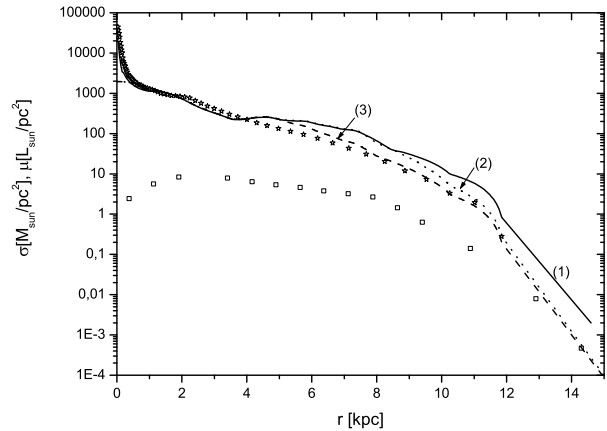}
\caption{\label{fig:ges} The solid line (labelled 1) shows the
surface mass density in the global disc model calculated based on
the rotation curve \citep{2011MNRAS.411...71H}. The star symbols
denote the surface brightness in the K filter
\citep{2003AJ....125..525J}. The squares denote the surface
density of HI+He based on \citep{1991AJ....101..456P}, extended by
the last three points, so that it converges to the dynamical mass
density. The dotted line (labelled 2) denotes the surface mass
density calculated based on the rotation curve 2 in
Fig.\ref{fig:rot}. The dashed line (labelled 3) denotes the
surface mass density calculated based on the rotation curve 3 in
Fig.\ref{fig:rot}. }
\end{figure}

\begin{figure} \centering
\includegraphics[width=0.5\textwidth]{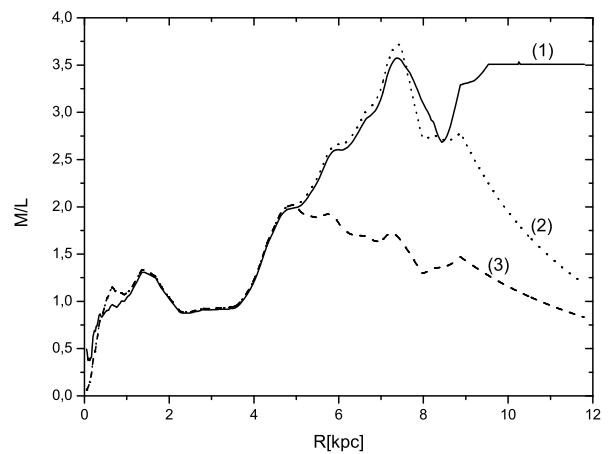}
\caption{\label{fig:mtol} The solid line (labelled 1) is the
mass-to-light ratio profile. The dotted line (labelled 2) is the
change in the mass-to-light ratio caused by a small modification
of the rotation velocity. The dashed line (labelled 3) is the
mass-to-light ratio profile corresponding to rotation curve 3 in
Fig.\ref{fig:rot}.}
\end{figure}

The resulting global surface mass density is shown in
Fig.\ref{fig:ges}
 and is compared with the surface brightness in the K
filter, using the surface density of the neutral hydrogen taken
from \citep{1991AJ....101..456P}. The corresponding (local)
mass-to-light ratio as a function of radial distance is shown in
Fig.\ref{fig:mtol}. 
This function grows, reaching $3.5$, then it
stabilizes at $r=9.6\,\kpc$ ((this plateau is a result of the
assumption made earlier). Although growing, the mass-to-light
ratio is low -- the ratio of the total mass to the total
brightness is $1.42$. \citet{2001ApJ...550..212B} stated that, for
spiral galaxies, the stellar mass-to-light ratio in near-infrared
is normally approximately $2$. Therefore, if we obtain results
that are, on average, close to this value, we conclude that most
of the galaxy mass consists of baryonic matter in the form of
stars. In conjunction with the declining rotation curve for large
radii, this fact allows us to conclude that the global disc model
is a good approximation of the distribution of matter in NGC 253.

An increase in the mass-to-light ratio is usually attributed to
the presence of a non-baryonic dark matter halo, which is expected
to manifest itself especially for large radii. We show that a
surface mass density can be obtained so that the corresponding
rotation curve of NGC 253 agrees with measurements within error
limits; the surface mass density gives a mass-to-light ratio that
decreases for large radii. To show this, the rotation curve
(labelled 1 in Fig.\ref{fig:rot}) has been modified slightly,
giving rise to the dotted curve labelled 2. Curves 1 and 2 differ
only slightly, and both agree with the measurements within error
limits.However, this small alternation of the rotation curve
results in a significant change in the mass-to-light ratio (see
Fig.\ref{fig:mtol}, dotted curve 2). It can be seen that the
mass-to-light ratio grows reaching $3.75$ at $r\approx7,\kpc$ and
then it decreases to $\approx1.25$. This example illustrates that
the mass-to-light ratio is very sensitive to modifications in the
rotation curve. It shows also that the rotation of the galaxy NGC
253 can be accounted for with a disc-like distribution of matter,
without the significant contribution of a spheroidal halo of dark
matter. Surely, this does not prove that nonbaryonic dark matter
is not present in NGC 253, but rather that, in some instances, its
influence might be significantly overestimated.

\section{Mass-to-light ratio and magnetic fields}
The magnetic fields present in NGC 253 are very strong, reaching
 $18 \muG$, and they influence the motion of ionized gas.
 Because the
rotation for large radii is measured using emission lines from a
diffused ionized gas \citep{2011MNRAS.411...71H}, the rotation
curve should be corrected for this non-gravitational interaction,
so that the dynamical mass can be correctly estimated. This effect
can reduce the mass-to-light ratio, and it should be especially
prominent for large radii
 \citep{2004AcPPB..35.2493K}.

To estimate it, first we assume a test rotation curve that agrees
with the measurements at small radii (within error limits) and
differs from them for radii greater than $\approx7\,\kpc$. The
test curve is shown in Fig.\ref{fig:rot} (dashed curve 3). We
assume that the test curve represents the rotation we would expect
to measure in the absence of magnetic fields (with the
gravitational field unchanged). Similar to
\citep{2012MNRAS.421.2155J}, we can now estimate the magnetic
field that is responsible for such a difference in rotation. We
begin with the Navier–Stokes equation
\begin{equation}\label{eq:navier}\br{\vec{v}\circ\vec{\nabla}}\vec{v}=
-\vec{\nabla}\Phi+\frac{1}{4\pi\,\rho}\br{\vec{\nabla}\times\vec{B}}\times
\vec{B}. \end{equation} Because we are interested in the rotation
in the plane z=0, we investigate only the radial component of this
equation. In the case of axial symmetry, assumed for the
simplicity of calculations, the derivatives with respect to $\phi$
are zero. Then, the radial part of equation \eqref{eq:navier}
reduces to $$-\frac{(\delta
v_{\varphi})^2}{r}=\frac{1}{4\pi\,\rho}\br{B_z\br{\partial_zB_r-\partial_rB_z}
-\frac{1}{r}B_{\phi}\partial_r\br{rB_{\phi}}}.$$ Here, $\br{\delta
v_{\varphi}}^2$ is the difference in the squares of rotation speed
for rotation curves 1 and 3 in Fig.\ref{fig:rot}, entirely due to
the magnetic field. In the case that we consider, we assume that
without the influence of magnetic fields, the rotation velocity
would be smaller, that is $\delta(v_\phi)^2=0$, whereas with
magnetic fields $\delta(v_\phi)^2>0$.

We analyse only the effect of the azimuthal component of the
magnetic field. The analysis of just a single component is
sufficient to demonstrate, to the order of a magnitude, the
possible effect of the influence of the magnetic field. From the
symmetry of the field with respect to the plane $z=0$, we can
expect the vertical component of the field to vanish on the disc.
If so (and from $\phi$-independence), the radial component of the
field will have had no effect on the rotation curve, because in
the radial part of $\br{\vec{\nabla}\times\vec{B}}\times \vec{B}$
the term containing the radial component of the magnetic field is
multiplied by the vertical component. Then, equation
\eqref{eq:navier} reduces to $(\delta
v_{\varphi})^2=\frac{1}{4\pi\rho}
B_{\varphi}\frac{\partial}{\partial r}(r B_{\varphi}) $  with the
solution \begin{equation}\label{eq:navier2}B_{\varphi}(r) =
\frac{1}{r} \sqrt{(r_{1})^2 (B_{\varphi}(r_{1}))^2  + 8 \pi
\int\limits_{r_1}^r \varrho (r) (\delta v_{\varphi}(\xi))^2  \xi
\ud{\xi}}. \end{equation}

\begin{figure} \centering
\includegraphics[width=0.5\textwidth]{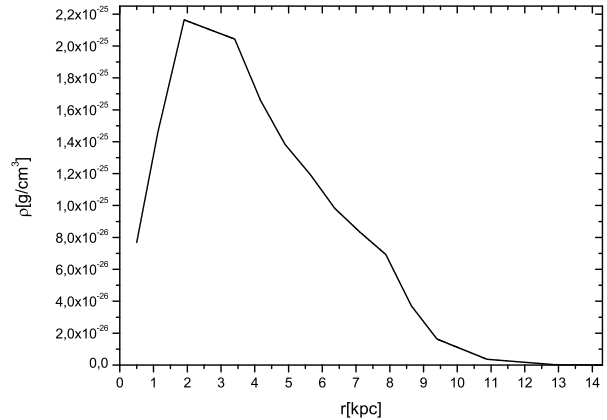}
\caption{\label{fig:gaz} Bulk density of HI in the disk plane
obtained from the surface density taken from
\citep{1991AJ....101..456P}} \end{figure}

\begin{figure}
   \centering
      \includegraphics[width=0.5\textwidth]{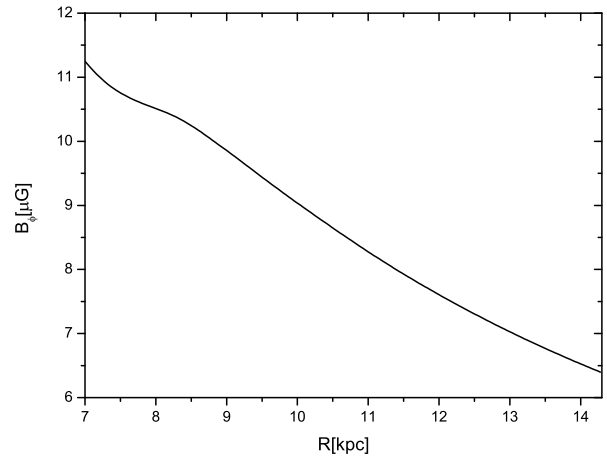}
      \caption{\label{fig:pole} Azimuthal magnetic field needed to change the rotation so that
the profile of the mass-to-light ratio is smaller and decreases
with the large galactic radius.}
       \end{figure}

To find particular solutions, we assume that $\rho$ decreases
exponentially with the altitude above the mid-plane and that its
column density is identical to the surface density of neutral
hydrogen from \citep{1991AJ....101..456P}, see Fig.\ref{fig:gaz}.
 These measurements are
available at radii smaller than those for the rotation. To
overcome this difficulty, we have extended the column density so
that the joint surface density of neutral hydrogen and helium
agrees for larger radii with the surface mass density obtained
based on rotation curve 3. The numerical solution to equations
\ref{eq:navier2} is shown in fig.\ref{fig:pole}.

The solid line represents the azimuthal magnetic field required
for the assumed difference between rotation curves 1 and 3  shown
in Fig.\ref{fig:rot}. The required field is no greater than $11.25
\muG$. For small radii, the field magnitude is larger; for larger
radii, it decreases with the radius.

The magnitude of the magnetic field in NGC 253 is of the order of
$7-18 \muG$ \citep{2009AN....330.1028H, 2005mpge.conf..156H}.It is
strongest in the central part of this galaxy, and it decreases in
the outer parts. At the radius of $\approx9 \kpc$, we note that
the magnetic field is $\approx10 \muG$
\citep{2005mpge.conf..156H}. We conclude that the magnetic field
in NGC 253 is sufficient to increase the rotation to the degree
assumed above. This change in the rotation diminishes the galaxy
mass by $18\%$ from $5.68\times10^{10}\msun$ (the mass in disk
model for rotation curve 1 inside $r<14.3\,\kpc$, the same value
as for curve 2) to $4.82\times10^{10}\msun$ (for the rotation
curve 3). The mass-to-light ratio was calculated based on the
rotation curve 3, and it is shown in Fig.\ref{fig:mtol} as the
dashed curve 3. The mass-to-light ratio has decreased
significantly; it reaches the maximal value of 2 and decreases for
$r> 5\,\kpc$. When accounting for the possible effect of the
magnetic field, the relative change that results for the estimated
galaxy mass is not large; however, the influence of the field
significantly changes the value and the behaviour of the
mass-to-light ratio for radii greater than $5\,\kpc$.

Of course, the situation that we consider is simplified. In NGC
253, there is also a small-scale turbulent field besides the
regular, large-scale magnetic field, and there are also other
phenomena that influence the magnetic field, such as gas flows out
of the disc plane or interactions with the radio halo. This makes
the real configuration of the magnetic field much more complex
than in the case we investigate. However, even such an idealized,
simplified situation allows us to demonstrate that the magnetic
fields can influence the motion of matter, and hence the rotation
curve of a spiral galaxy.

\section{Summary}
The latest measurements show that the rotation curve of NGC 253
decreases for large radii, which makes this galaxy a natural
candidate for a disc-like galaxy without (or with only a small
fraction of) non-baryonic dark matter. Indeed, the mass
distribution can be consistently described in the global disc
model, without introducing any massive spherical component. The
resulting mass-to-light ratio (in the K filter) is low. In
addition, this ratio can be reduced further by the very strong
magnetic field present in NGC 253, which should influence the
motion of ionized gas. Recall that the mass-to-light ratio was too
high in the maximal disc model of NGC 253 when the surface
brightness in the B filter was used \citep{2011MNRAS.411...71H}.
Moreover, in contrast to \citet{2011MNRAS.411...71H}, we do not
assume in advance that this ratio is a constant. In our approach,
the mass-to-light ratio is a local quantity, a function of the
radius.

With the rising rotation curve of \citet{1991AJ....101..456P} the
mass-to-light ratio increases, reaching 4 in the region where the
measurements of rotation are available. When the possible
influence of magnetic fields is taken into account, the ratio is
reduced, reaching the maximal value of $2.25$. The required field
decreases with the radius, starting from $11.5 \muG$. With this
field, the mass-to-light ratio increases with the radius, and thus
it is reduced differently than in the case with the decreasing
rotation curve.

The fall-off in the rotation curve for large radii suggests that
NGC 253 is a disc-like object, without a significant amount of
non-baryonic dark matter. However, even the analysis based on the
increasing rotation curve has given a low mass-to-light ratio.
Additionally, this ratio is reduced when the possible influence of
the magnetic field is taken into account. Recall that in
calculating the mass-to-light ratio, the surface mass density of
gas was not subtracted from the total surface mass density,
because the measurement of neutral hydrogen ended earlier than the
measurements of surface brightness and of rotation. Therefore, in
this work we have obtained an upper bound for the mass-to-light
ratio (there are other mass components apart from stars that
contribute to the surface mass density).

Our analyses of NGC 253, as well as a previous analysis of NGC
891, show that values of the mass-to-light ratio calculated in the
disc model, especially for large radii, are very sensitive to even
small changes in the rotation curve, and they might also be
influenced by other factors, such as the magnetic field, which
have not yet been taken into account. Using estimates of the order
of a magnitude, essentially, we have shown that a large-scale
field of about the observed strength can influence the rotation
curve. In certain galaxies, this might significantly affect
deductions on the size of the dark matter halo, which are based on
the rotation curve. Therefore, we should be cautious in drawing
conclusions about the dark matter abundance in spiral galaxies
from the increase in their mass-to-light ratio, if we do not know
and cannot account for the various factors that can influence this
ratio.

\section{Acknowledgments}
We thank Reiner Beck from the Max Planck Institute f\"{u}r
Radioastronomie for turning our attention to NGC253.

\bibliography{NGC253}
\bibliographystyle{mn2e}
\end{document}